\documentclass[12pt]{article}
\usepackage[cp1251]{inputenc}
\usepackage[english]{babel}
\usepackage{epsfig}
\usepackage{amssymb}
\usepackage{amsmath}
 0

\begin{document}

\title{Proper orthogonal decomposition of eigen modes in a gas affected by a mass force}
\author{Sergey Leble, Anna Perelomova
\small\\  Gdansk University of Technology, \\Faculty of Applied
Physics and Mathematics,\\ ul. Narutowicza 11/12, 80-233 Gdansk,
Poland,
\small \\ leble@mif.pg.gda.pl,\;anpe@mif.pg.gda.pl \\  \\[2ex] }
\maketitle

\renewcommand{\abstractname}{\small Summary}
\begin{abstract}
The relations connecting perturbations in acoustic and entropy modes in a gas affected by a constant mass force, are derived. The background temperature of a gas may vary in the direction of an external mass force. The relations are independent on time. They make possible to decompose the total vector of perturbations into acoustic and non-acoustic parts uniquely at any instant.  
In order to do this, three quantities are required, according to the number of modes. In one dimension, the reference quantities may be total perturbations in entropy, pressure and velocity. The total energy of  flow is determined. The examples of dismemberment of the total field into acoustic and entropy parts relate to the unperturbed temperature of a gas which linearly depends on the spacial co-ordinate.
\end{abstract}

{\bf Key words:}Atmospheric perturbations, Initialization of hydrodynamic field, Non-uniform media, Sound propagation, Heating.  \\
{\bf PACS No.} 43.25.Cb, 43.28.Bj

\section{Introduction}
In experiments  of 1994 (Transarctic Acoustic Propagation Experiment, TAP), coherent acoustic transmissions via Arctic basin were studied. It was designed to monitor changes in Arctic Ocean temperature and    sea ice properties.  TAP demonstrated that the low frequency (19.6-Hz) signals propagated with both sufficiently low loss and high phase stability to support the coherent pulse compression processing  and the phase detection of the signals. These yield time delay measurements an order of magnitude better than what is required to detect the  year changes in travel time caused by interannual and longer term changes in Arctic Ocean temperature. The TAP data provided propagation loss measurements to compare with the models to be used for correlating modal scattering losses with sea ice properties for ice monitoring. The travel times  measurements indicated a temperature changes   in the Arctic atmosphere, which has been confirmed by direct measurement from icebreakers and submarines, demonstrating the utility of acoustic thermometry \cite{MGB}. Climatological atmospheric velocity models predict   infrasound signals from sources that occur at mid-northern latitudes. Such infrasound data have been used to locate a bolid explosions in space and time.  In \cite{HDW} authors  analyze travel time picks and use 3-D ray tracing to generate synthetic travel times based on various atmospheric models to show that the seismic network data instead reveal a predominant  propagation direction. A sudden stratospheric warming event that reversed the zonal wind flow explains   propagation properties.

A time-dependent, nonlinear, fully compressible, axisymmetric, f-plane, numerical model is used to simulate the propagation of acoustic waves in the mesosphere and thermosphere by intense deep convection in the troposphere \cite{WSB}. The simulations show that major convective storms in the tropics launch acoustic waves into the mesosphere-thermosphere directly above the storm centers . The principal feature of the overhead acoustic wave field in the period interval of 3 to 5 min is a trapped oscillation below about 80 km altitude with a period of 5 min and a nearly vertically propagating wave with about a 3-min period above this height. The vertically propagating thermospheric acoustic oscillations are waves propagating upward from the thunderstorm source through the stratosphere-mesosphere. These predominantly 3-min waves are strongly driven for about 30 min after the storm event and weaken with time thereafter. The vertically propagating thermospheric acoustic waves may be the source of the F-region 3-min oscillations. Intense acoustic disturbances directly above thunderstorms may also be responsible for localized heating of the thermosphere \cite{P1}.

There is an important ingredient on which a direct problem of sound propagation or inverse problems (such as acoustic thermometry) are based. It is a necessity the proper division of an atmospheric disturbance onto the characteristic types and hence a possibility to formulate a problem of acoustic wave generation propagation and reflection as one of them.
The theoretical and numerical models as e.g. of \cite{WSB} describing dynamics of gases and liquids affected by external forces are of great interest in geophysics, meteorology and wave theory \cite{Ped}. The authors believe that the theoretical models are more desirable than the numerical methods, which are time-consumed and require large computer power and special converging and stable numerical methods. Numerical methods, if necessary, are much more economical and successful, if they rely on the reasonable theoretical models. Moreover, some important problems may be resolved only theoretically. Some of these problems are considered in this study. Among other, they are: how many types of fluid motion exist, what they are and what energy associates with every type of motion. The external forces make the background of
a fluid non-uniform, with background density, temperature and pressure
depending on spatial coordinates. That essentially complicates the definition
of linear modes (motions of infinitely small amplitude) taking
place in such non-uniform media \cite{Le}. The number of roots of dispersion
equation, if it is possible to determine them, or branches of possible types of motion, equals
number of the conservation equations \cite{Brech},\cite{Ped}. Even in the simplest case of flow in one dimension,
the dispersion relations may be introduced over all wave-length range only if the background pressure and density depend on the coordinate exponentially [1]. Anyway, there are three
types of motion in one dimension: two acoustic branches and, if the thermal conduction of a fluid is
ignored, the stationary (entropy, non-wave) mode. In the flows exceeding one dimension, the buoyancy waves appear and interact \cite{Ped},\cite{Le}.

In this study, modes of one-dimensional flow are determined by independent on time relations linking hydrodynamic perturbations. They are fixed for any mode and valid for arbitrary dependence of the background temperature on co-ordinate.  These relations give possibility to distinguish modes analytically at any instant, to conclude about energy of everyone from them 
and to predict their dynamics.  We will consider volumes of an ideal gas affected by a constant mass force, with variable temperature of the background.  The first results allowing to distinguish modes due to relations of specific perturbations, were obtained relatively to the motion of atmospheric gas affected by gravity in \cite{Le},\cite{BrLePe}. Fluids different from ideal gases, including liquids, are briefly discussed in Concluding Remarks.
It is undoubtedly of importance in applications of meteorology and atmospheric dynamics. The case of an ideal gas with constant temperature is considered in Sec.3. The examples relating to the linear dependence of the background temperature on vertical co-ordinate in the field of constant gravity force, are discussed in Sec.4.
\section{Statement of problem}
\subsection{Conservation equations and the total energy}

The equations governing fluid in absence of the first, second viscosity and thermal conduction manifest
conservation of momentum, energy and mass.  They 
are generally nonlinear. We start from the linearized differential equations in terms of variations of pressure and density, $p'$ and $\rho'$ from hydrodynamically stable stationary functions $\overline p$, $\overline \rho$, which are not longer constants:
$$\frac{\partial \overrightarrow V }{\partial t}=-\frac{\overrightarrow\nabla p'}{\overline \rho}
- \overrightarrow g,$$
$$\frac{\partial  p' }{\partial t}=-\overrightarrow V\cdot\left(\overrightarrow \nabla \overline p\right)-\gamma\overline\rho\left(\overrightarrow\nabla\cdot \overrightarrow V\right) , \eqno(1)$$
$$\frac{\partial\rho'}{\partial t}=-\overrightarrow V\cdot\left(\overrightarrow \nabla \overline \rho\right)-\overline\rho\left(\overrightarrow\nabla\cdot \overrightarrow V\right).$$
The mean flow is absent, so that its velocity equals zero, $\overline{\overrightarrow V(x,y,z)}\equiv \vec 0$.
The external force will be described by the gravity acceleration $\overrightarrow g=(0,0,g)$, though it may involve readily other mass forces including non-inertial ones. The flow of an ideal gas is considered, which internal energy $e$ in terms of pressure and density takes the form
$$e=\frac{p}{(\gamma-1)\rho},\eqno(2)$$
where $\gamma=C_p/C_v$ denotes the specific heats ratio. Eqs (1) 
describe gas motion of infinitely small amplitude. The background pressure and density follow from the zero order stationary equality,
$$\frac{d\overline p(z)}{d z}=-g\overline \rho(z).\eqno(3)$$
The background quantities supporting the equilibrium distribution of temperature $T_0(z)$, take the form
$$\overline \rho(z)=\frac{\overline \rho(0)H(0)}{H(z)}\exp\left(-\int_0^z\frac{dz'}{H(z')}\right),\quad H(z)=\frac{T_0(z)(C_p-C_v)}{g}. \eqno(4)$$
It is well-established, that distribution of the background temperature is equilibrium if the parameter of static stability $\nu(z)$ is positive \cite{Ped},
$$\nu(z)=\gamma-1+\gamma\frac{dH(z)}{dz}>0.\eqno(5)$$
For some reason, it is convenient to introduce the quantity $\varphi'$ instead of perturbation in density, 
$$\varphi'=p'-\gamma\frac{\overline p}{\overline \rho}\rho'.\eqno(6)$$
The integral
$$\varepsilon=\frac{1}{2}\int\left(\overline \rho\overrightarrow V^2+\frac{p'^2}{\gamma \overline p}+\frac{\varphi'^2}{\gamma\nu(z)\overline p}\right) dv \eqno(7)$$
is invariant, where
$$v=\left\{-\infty<x,y<\infty,\;0\leq z\leq h\right\},\eqno(8)$$
and $h$ may be infinity. It readily follows from Eqs (1)--(7), that
$$\frac{\partial \varepsilon}{\partial t}=-\int \overrightarrow{\nabla}\cdot(p'\overrightarrow V)dv =-\oint_{\sigma(v)}p'\overrightarrow Vd\vec\sigma=0,\eqno(9)$$
where $\sigma$ is a surface circumscriptive the volume $v$. The invariance of $\varepsilon$ manifests  conservation of the total energy of gas, which includes kinetic, barotropic and thermal parts.
For $\varepsilon$ to be invariant, there is a certain freedom to establish the  boundary conditions at $z=0$ and $z=h$:
$V_z(z=0)=V_z(z=h)=0$, $p'$ is any smooth function (impermeability condition across the boundaries) , or, for example, $V_z(z=0)=0,\;p'(z=h)=0$.

\section {Spectral problem and definition of modes}
Introducing the new set of variables,
$$P=p'\cdot\exp\left(\int_0^z \frac{dz'}{2H(z')}\right),\;\Phi=\varphi'\cdot\exp\left(\int_0^z \frac{dz'}{2H(z')}\right),\;\vec U=\vec V\cdot\exp\left(-\int_0^z \frac{dz'}{2H(z')}\right),\eqno(10) $$
one may readily rearrange Eqs (1) into the following set:
$$\frac{\partial U_x }{\partial t}=-\frac{1}{\overline\rho(0)}\frac{\partial P}{\partial x},$$
$$\frac{\partial U_y }{\partial t}=-\frac{1}{\overline\rho(0)}\frac{\partial P}{\partial y},\eqno(11)$$
$$\frac{\partial U_z }{\partial t}=\frac{1}{\overline\rho(0)}\left(\frac{\gamma-2}{2\gamma H(0)}-\eta(z)\frac{\partial}{\partial z}\right)P+\frac{\Phi}{\gamma H(0)\overline\rho(0)},$$
$$\frac{\partial P }{\partial t}=-\gamma g H(0)\overline\rho(0)\left(\frac{\partial U_x }{\partial x}+\frac{\partial U_y }{\partial y}+\frac{\partial U_z }{\partial z}\right)-g\overline\rho(0)\frac{\gamma-2}{2\eta(z)}U_z, $$
$$\frac{\partial\Phi}{\partial t}=-\frac{\nu(z)}{\eta(z)}g\overline\rho(0)U_z,$$
where $\eta(z)=\frac{H(z)}{H(0)}$. Eqs (11) determine the spectral problem.

\subsection{Case of the constant background temperature}
The analytical analysis of the dispersion relations and determined by them modes may be proceeded in the case of constant $T_0$ and hence constant $H$. In this case, $\eta=1$ and $\nu=\gamma-1$. The following matrix equality represents the system (11),
$$\frac{\partial}{\partial t}\Psi(\vec r,t)=\mathbf L\left(\frac{\partial}{\partial x},\frac{\partial}{\partial y},\frac{\partial}{\partial z}\right)\Psi(\vec r,t),\eqno(12)$$
where
$$\Psi=[\;U_x\;\;\;U_y\;\;\;U_z\;\;\;P\;\;\;\Phi\;]^T,\eqno(13)$$
$\vec r=(x,y,z)$ and $\mathbf L$ is the matrix operator depending on spacial partial derivatives.

The condition of algebraic solvability of Eqs (12) may be established by the Fourier transformation using the basis functions $\exp( i k_x x + i k_y y+i k_z z)$, $\Psi(\vec r,t)=\int_{-\infty}^{\infty}\exp(-i\omega t+i k_x x + i k_y y+i k_z z)\psi(k_x,k_y,k_z)dk_xdk_ydk_z + cc$,
$$Det||i\omega\mathbf{I}-\mathbf{l}(k_x,k_y,k_z)||=0,\eqno(14)$$
where $\bf{I}$ is the unit matrix, $\mathbf l$ represents the matrix operator $\mathbf L$ in the space of Fourier transforms. The dispersion equation (14) determines five roots of dispersion equation, or dispersion relations, everyone responsible for an especial type of gas motion. There are four wave modes, denoted by indices $1,2,3,4$, and the entropy mode, marked by $0$:
$$\omega_0=0,\eqno(15)$$
$$\omega_{1,2}=\pm\sqrt{\frac{\gamma g H}{2}}\sqrt{k_x^2+k_y^2+k_z^2+\frac{1}{4H^2}+\sqrt{\left(k_x^2+k_y^2+k_z^2+\frac{1}{4H^2}\right)^2-\frac{4(\gamma-1)}{\gamma^2H^2}(k_x^2+k_y^2)}},$$
$$\omega_{3,4}=\pm\sqrt{\frac{\gamma g H}{2}}\sqrt{k_x^2+k_y^2+k_z^2+\frac{1}{4H^2}-\sqrt{\left(k_x^2+k_y^2+k_z^2+\frac{1}{4H^2}\right)^2-\frac{4(\gamma-1)}{\gamma^2H^2}(k_x^2+k_y^2)}}.$$
Vectors of perturbations $\Psi_n$ ($n=0,\dots 4$) correspondent to eigenvalues $i\omega_n$, form the Hilbert space $L^2(v)$ with the scalar product
$$\left\langle \Psi_n,\Psi_m\right\rangle=\int  \left(\rho_0\overrightarrow U_n\cdot\overrightarrow U_m+\frac{P_n P_m}{\gamma g H\rho_0}+\frac{\Phi_n\Phi_m}{\gamma(\gamma-1)g H \rho_0}\right)dv\eqno(16)$$
and the invariant
$$E=\frac{1}{2}\int \left(\rho_0|\overrightarrow U|^2+\frac{P^2}{\gamma g H\rho_0}+\frac{\Phi^2}{\gamma(\gamma-1)g H \rho_0}\right)dv,\eqno(17)$$
where $\overrightarrow U$, $P$ and $\Phi$ represent a sum of all parts of the eigenvectors,
$$\overrightarrow U=\sum_{n=0}^4\overrightarrow U_n,\;P=\sum_{n=0}^4P_n,\;\Phi=\sum_{n=0}^4\Phi_n.\eqno(18)$$
The set $\Psi_n$ 
form complete set of eigenvectors. That is true for the self-adjoint boundary conditions. It may be readily established, that $iL$ is symmetric in $L^2(v)$ ($0\leq n,m\leq 4$):
$$\left\langle i L\Psi_n,\Psi_m \right\rangle-\left\langle \Psi_m,i L\Psi_n \right\rangle=
i\oint_{\sigma(v)}\left(P_n\overrightarrow U_m+P_m\overrightarrow U_n\right)d\vec\sigma=0.\eqno(19)$$
The most important physically condition of impermeability at the upper and lower boundaries, $z=0$ and $z=h$, is self-adjoint: $U_z(z=0)=U_z(z=h)=0$. It is follow from Eqs (16)--(18) that $E$ may be decomposed in to specific energies, $E=\sum_{n=0}^4E_n$.

\section{Decomposition of
 acoustic and entropy modes in one-dimensional flow. General case}
In one dimension ($k_x=k_y\equiv 0$) and constant $T_0$, the dispersion relations (15) determine three modes, or, in the other words, possible motions in the gas. Two of them are acoustic, describing different direction of sound propagation, and the last one is stationary, or entropy mode. It is responsible for stationary variations in pressure and density which result in stationary change in the gas temperature. In the absence of mass force, this mode would be isobaric. In the case of variable $T_0$, the dispersion equation, valid over all wave-length range, can not longer be written on, but the modes may be determined by relations linking field perturbations in this case as well. The completeness of the set of eigenvectors allow to represent the total vector of perturbations as a sum of acoustic and entropy vectors at any instant,
$$\Psi(z,t)=\sum_{n=1}^3\Psi_n(z,t)=\Psi_0(z,t)+\Psi_a(z,t), \eqno(20)$$
where index $a$ denotes summary acoustic vector.
Eqs (11) yield the relation connecting $P(z,t)$ and $\Phi(z,t)$ in both acoustic branches,
$$P_a=\frac{\eta(z)}{\nu(z)}\left(\frac{\gamma-2}{2\eta(z)}+\gamma H(0)\frac{\partial}{\partial z}\right)\Phi_a,\eqno(21)$$
and the link for the
stationary entropy mode,
$$\Phi_0=\left(-\frac{\gamma-2}{2}+\gamma H(0)\eta(z)\frac{\partial}{\partial z}\right)P_0.\eqno(22)$$
The scalar product of eigenvectors is given by equality
$$\left\langle \Psi_n,\Psi_m\right\rangle=\int  \left(\overline \rho \overrightarrow U_n\cdot\overrightarrow U_m+\frac{P_n P_m}{\overline p}+\frac{\Phi_n\Phi_m}{\nu(z)\overline p}\right)dv.\eqno(23)$$
The total vector of perturbations is a sum of orthogonal modes
$$\Psi(z,t)=\left[\begin{array}{c}U_z\\P\\ \Phi\end{array}\right]=\Psi_1(z,t)+\Psi_2(z,t)+\Psi_3(z,t)\equiv\Psi_a(z,t)+\Psi_0(z,t)=\eqno(24)$$ $$
\left[\begin{array}{c}U_{a,z}\\\left(\frac{\gamma-2}{2\eta(z)}+\gamma H(0)\frac{\partial}{\partial z}\right)\frac{\eta(z)}{\nu(z)}\Phi_a\\ \Phi_a\end{array}\right]+\left[\begin{array}{c}0\\P_0\\\left(-\frac{\gamma-2}{2}+\gamma H(0)\eta(z)\frac{\partial}{\partial z}\right)P_0\end{array}\right].$$
Excluding $P_0$ from the system (24), one obtains the equation describing $R(z,t)=\frac{\eta(z)}{\nu(z)}\Phi_a(z,t)$:
$$\left(1+2H(0)\frac{d\eta(z)}{dz}-4H^2(0)\eta^2(z)\frac{\partial^2}{\partial z^2}\right)R(z,t)=\frac{2\eta(z)}{\gamma^2}\left(2\Phi +\left(\gamma-2-2\gamma H(0)\eta(z) \frac{\partial }{\partial z}\right)P\right).\eqno(25)$$
Its solution consists of that of the homogeneous equation and a partial solution,
$$R(z,t)=R_2\int_0^zR_1(z')D(z',t)dz'-R_1\int_0^zR_2(z')D(z',t)dz'+C_1R_1+C_2R_2,\eqno(26)$$
where $R_1(z)$, $R_2(z)$ are determined by
$$R_1(z)=\exp\left(-\int_0^z\frac{dz'}{2H_0\eta(z')}\right),\quad R_2(z)=R_1\int_0^z\frac{dz'}{R_1^2(z')},\eqno(27)$$
and
$$D(z,t)=-\frac{1}{2H^2(0)\eta(z)\gamma^2}\left(2\Phi +\left(\gamma-2-2\gamma H(0)\eta(z) \frac{\partial }{\partial z}\right)P\right).\eqno(28)$$
$R(z,t)$ uniquely determines $P_a$ and stationary quantities $\Phi_0=\Phi-\Phi_a$ and $P_0$ in accordance to Eqs (24).

\section{Linear dependence of $H$ on $z$}
\subsection{Explicit formula choice}
In this case, $\eta=1+\alpha z$, $\alpha$ is some  non-zero constant, and $\nu=1-\gamma+\gamma\alpha H(0)$, and $R_1$, $R_2$ take the following form
$$R_1(z)=\left(\frac{1}{1+\alpha z}\right)^{1/2\alpha H(0)},\quad R_2(z)=H(0)\left(\frac{1}{1+\alpha z}\right)^{1/2\alpha H(0)}\frac{-1 + (1 + \alpha z)^{1 + 1/\alpha H(0)}}{1 + \alpha H(0)
}.\eqno(29)$$
Eqs (25)--(29) allow to distinguish uniquely acoustic and entropy mode in any vector of total perturbations.
Some simple conclusions follow immediately from Eqs (24).

\subsection{Contribution of only entropy mode in the total perturbation}
In this case,
$$\Phi\equiv\Phi_0=\left(-\frac{\gamma-2}{2}+\gamma H(0)(1+\alpha z)\frac{\partial}{\partial z}\right)P.\eqno(30)$$
In illustrations, we take $P\equiv P_0$ in the form of a Gaussian impulse (a) and its derivative (b),
$$(a)\;P=\pi\exp(-(z-z_0)^2/\beta^2 H(0)^2),\;(b)\;P=-2\pi\frac{\exp(-(z-z_0)^2/\beta^2 H(0)^2)(z-z_0)}{H(0)\beta^2},\eqno(31)$$
where $\beta, \pi$ denote the characteristic width of the impulse in units $H$ and its amplitude. It would be superfluous to mention, that relations between field perturbations specifying every mode, Eqs (24), are valid at any instant. So, in this subsection and two subsections below, we do not determine the time which the samples of perturbations relate to. For definiteness, $z_0=3H(0),\;\beta=0.3$, and $\gamma=1.4$.
\begin{center}
\epsfig{file=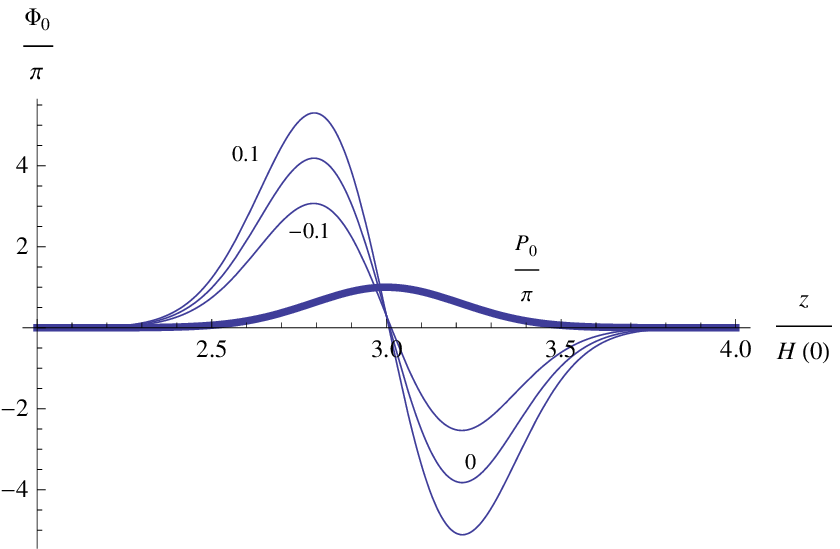, height=5cm,width=6cm,clip=,angle=0}
\epsfig{file=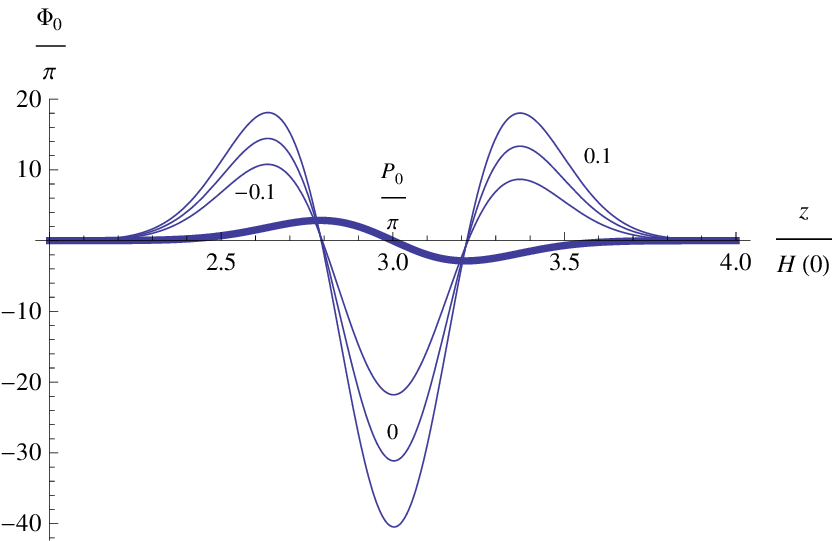, height=5cm,width=6cm,clip=,angle=0}
\end{center}
\begin{center}\textbf{a\hspace{6cm}b}\end{center}
\begin{center}\textbf{Fig.1} Excess pressure and entropy in the entropy mode for different $\alpha H(0)$ ($-0.1,0,0.1$).\end{center}

\subsection{Contribution of only sound in the total perturbation}
In accordance to Eqs (24),
$$P\equiv P_a=\left(\frac{\gamma-2}{2(1+\alpha z)}+\gamma H(0)\frac{\partial}{\partial z}\right)\frac{1+\alpha z}{1-\gamma+\gamma H(0)\alpha}\Phi.\eqno(32)$$ $\Phi\equiv \Phi_a$ is a Gaussian pulse and its derivative,
$$(a)\;\Phi=\pi\exp(-(z-z_0)^2/\beta^2 H(0)^2),\;(b)\;\Phi=-2\pi\frac{\exp(-(z-z_0)^2/\beta^2 H(0)^2)(z-z_0)}{H(0)\beta^2},\eqno(33)$$
The values of $z_0$, $\beta$, and $\gamma$ are the same as in the above subsection.
\begin{center}
\epsfig{file=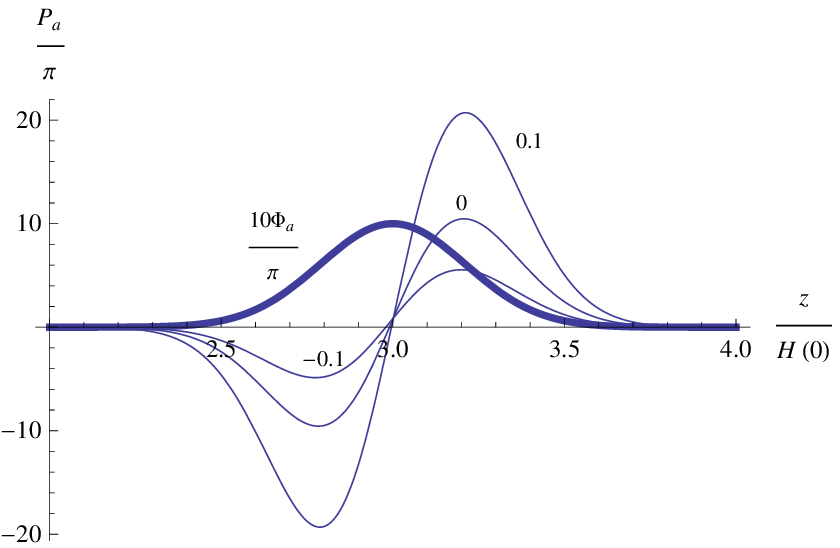, height=5cm,width=6cm,clip=,angle=0}
\epsfig{file=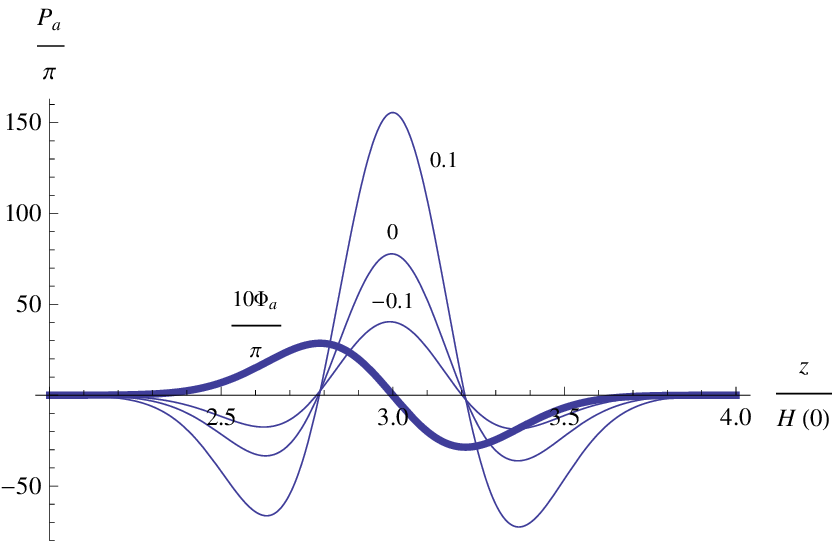, height=5cm,width=6cm,clip=,angle=0}
\end{center}
\begin{center}\textbf{a\hspace{6cm}b}\end{center}
\begin{center}\textbf{Fig.2} Excess pressure and entropy in the sound for different $\alpha H(0)$ ($-0.1,0,0.1$).\end{center}

\subsection{Zero total entropy}
The specific perturbations in pressure for the stationary mode is taken in both forms (a,b) as in Eq.(31), and the correspondent entropy is given by Eq.(30). Perturbation in entropy for sound is $\Phi_a=-\Phi_0$, and the excess pressure relates to $\Phi_a$ by means of Eq.(32). Fig.3 shows excess pressure for acoustic and entropy modes in the cases (a) and (b) for different $\alpha$.

\begin{center}
\epsfig{file=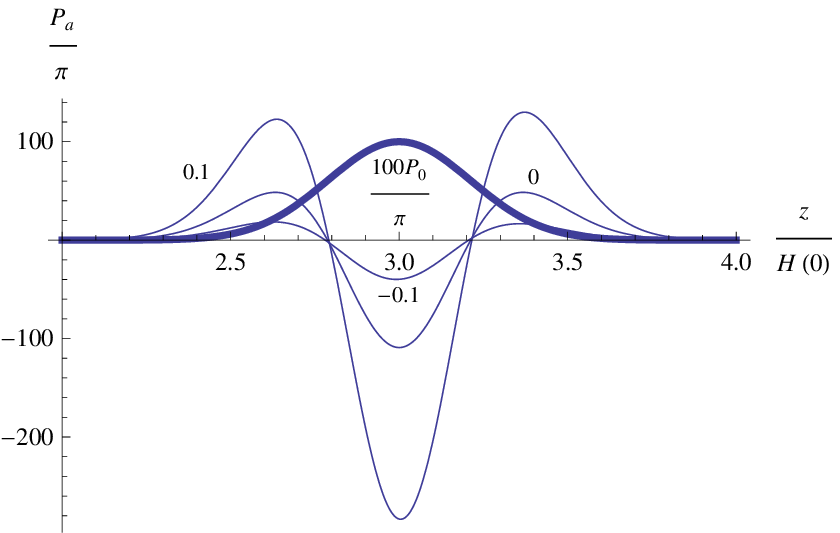, height=5cm,width=6cm,clip=,angle=0}
\epsfig{file=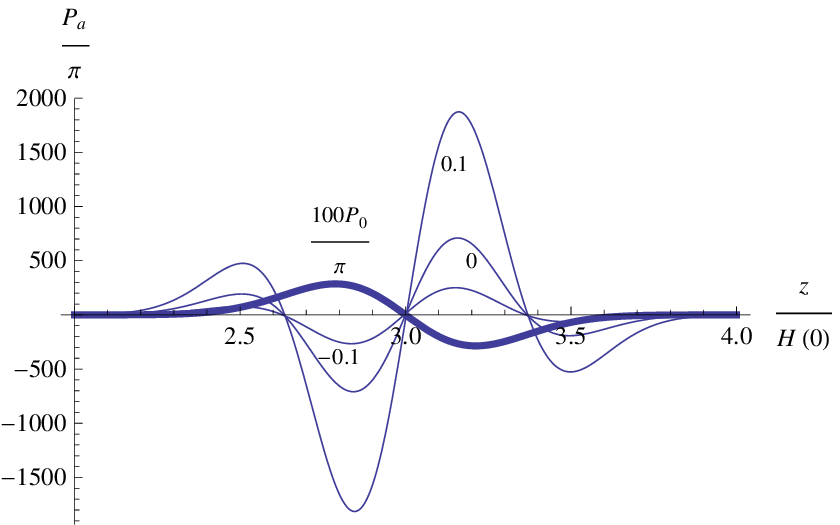, height=5cm,width=6cm,clip=,angle=0}
\end{center}
\begin{center}\textbf{a\hspace{6cm}b}\end{center}
\begin{center}\textbf{Fig.3} Case of zero total entropy. $\alpha H(0)$ takes values ($-0.1,0,0.1$).\end{center}

\section{Concluding remarks}
The choice of reference quantity in sound ($\Phi_a$) was made due to simplicity of relation linking $P_a$ and $\Phi_a$ for both acoustic branches. Note that in the limit of homogeneous gas ($g\rightarrow 0$, $H\rightarrow\infty$, but the product $gH(0)$ remains constant), $\gamma g H(0)$ is the squared sound of velocity in the uniform ideal gas (about the limit look the numerical investigation \cite{KL}). In this case, $\varphi'$ means the quantity proportional to perturbation in the entropy, $\varphi'=(\gamma-1)\overline \rho T_0 s'$. It is identically zero in both acoustic branches and is not longer suitable to be a reference quantity in them. Instead of, perturbation in density or in temperature may be chosen. In order to conclude about velocity of the acoustic mode, the knowledge of relation linking it with $\Phi_a$ is required. The relations differ in sign for different acoustic modes. That follows from the conservation system (11): $U_{1,z}(z,t)=K\Phi_1(z,t)$, $U_{2,z}(z,t)=-K\Phi_2(z,t)$, where $K$ is some integro-differential operator. 
It may be concluded from these equalities and Eqs (24), that if  $\Phi_{1}(z,t)=-\Phi_{2}(z,t)$ and $P_{1}(z,t)=-P_{2}(z,t)$,
the total field is represented by the entropy mode and acoustic field with non-zero velocity
$U_{z}(z,t)=U_{1,z}(z,t)+U_{2,z}(z,t)\equiv U_{1,z}(z,t)\equiv U_{2,z}(z,t)$ (hence, in general, the non-zero kinetic energy). Zero total kinetic energy
means inevitably equal perturbations in pressure and entropy of both acoustic branches, $P_{1}(z,t)=P_{2}(z,t)$, $\Phi_{1}(z,t)=\Phi_{2}(z,t)$.
The relations (24) are valid \textit{at any instant}. In the other words, though perturbations in every quantity varies with time, links inside every mode remain independent on time.

To make approximate evaluations of $K$, one may consider the waveforms which spectrum includes mostly large wavenumbers $k_z$, $k_zH(0)>>1$ \cite{Jon}. That allows to expand acoustic eigenvalues in the Taylor series and do obtain finally relations for every acoustic branch, propagating over the isothermal gas:
$$U_{1,z}(z,t)=-\frac{g\gamma^2}{8\rho_0(\gamma-1)(\gamma g H(0))^{3/2}}\int_0^z \Phi_{1}(z',t)dz',\eqno(34)$$
$$ U_{2,z}(z,t)=\frac{g\gamma^2}{8\rho_0(\gamma-1)(\gamma g H(0))^{3/2}}\int_0^z \Phi_{2}(z',t)dz'.$$
The simple conclusion from Eqs (34) is that if $\int_0^z (\Phi_{1}(z',t)+\Phi_{2}(z',t))dz'$, the kinetic energy at this instant is zero.

The relations linking  $U_z$ with $P$ and $\Phi$ in a fluid affected by a mass force, are integro-differential. The exact links of excess pressure, density and velocity in unbounded volumes of gas with constant $H$, are derived exactly with regard to one-dimensional flow by one of the authors in \cite{P1,P2,P3}. They are valid for an ideal gas or any other fluid. For fluid different from an ideal gas, the equation of state, Eq.(2) should be corrected \cite{P2}. That corrects also dynamic equation governing an excess pressure, the second one from Eqs (1), and therefore, definition of modes and even definition of $P$,$\Phi$ and $\overrightarrow U$ (Eqs (10)).  It was shown in the study \cite{P2}, that in the case of constant $H\equiv H(0)$, the relations (10) should be completed by the factor $\exp(\xi z)$, where
$$\xi=-\frac{AgH+B}{2H(gH-B)},\eqno(35)$$
where $A$ and $B$ follow from the leading-order series of excess internal energy,
$$e'=A\frac{p'}{\overline \rho}+B\frac{p'}{\overline \rho}.\eqno(36)$$
$\xi$ equals zero only in an ideal gas. Estimations for water at normal conditions result in $\xi\approx 0.85/2H$. So that, even in a linear fluid flow over the background of constant temperature, the results can not be generalized by common replacing $\gamma$ by $c^2\overline\rho(0)/\overline p(0)$, where $c$ denotes the sound velocity over the fluid without mass force, and $\overline p$ denotes the unperturbed internal pressure in it.
 
Illustrations of Sec.5 relate to $\alpha H(0)$ taking values $-0.1$, $0$ or $0.1$. In the model of standard atmosphere, there are some extended domains of almost linear dependence of $H$ on $z$: $\alpha H(0)$ is about $-0.2$ over the domain of $z$ between $0$ and $10$ kilometers, $\alpha H(10)\approx0$ between $10$ and $20$ kilometers and $\alpha H(30)\approx0.1$ between $30$ and $45$ kilometers. These quantities may considerably vary depending on season, daytime and meteorologic conditions.

\end{document}